\documentclass[12pt,preprint]{aastex}

\begin{document}

\title{Rapid Formation of Super-Earths around M Dwarf Stars}

\author{Alan P.~Boss}
\affil{Department of Terrestrial Magnetism, Carnegie Institution of
Washington, 5241 Broad Branch Road, NW, Washington, DC 20015-1305}
\authoremail{boss@dtm.ciw.edu}

\begin{abstract}

While the recent microlensing discoveries of super-Earths orbiting two
M dwarf stars have been taken as support for the core accretion mechanism
of giant planet formation, we show here that these planets could
also have been formed by the competing mechanism of disk instability,
coupled with photoevaporative loss of their gaseous envelopes by a 
strong external source of UV radiation, i.e., an O star. M dwarfs that 
form in regions of future high-mass star formation would then be expected 
to have super-Earths orbiting at distances of several AU and beyond,
while those that form in regions of low-mass star formation would
be expected to have gas giants at those distances. Given that most
stars are born in the former rather than in the latter regions, M 
dwarfs should have significantly more super-Earths than gas giants, 
as seems to be indicated by the microlensing surveys.

\end{abstract}


\section{Introduction}

Microlensing surveys have discovered recently two ``super-Earths''
orbiting M dwarf stars, planets with masses of $\sim 5.5 M_\oplus$
(Beaulieu et al. 2006) and $\sim 13 M_\oplus$ (Gould et al. 2006).
As microlensing detections yield only the ratio of the lensing planet mass 
to the lensing star mass, these planetary mass estimates rely heavily 
on the unproven assumption that the lensing star is an M dwarf. The 
super-Earths have been interpreted as being ``failed cores'' produced
by the first step of the core accretion mechanism for gas giant
planet formation. The super-Earths presumably failed to become
gas giant planets because the growth of solid cores by collisional
accumulation proceeds considerably slower at a fixed orbital radius 
around an M dwarf star than around a G dwarf star (Laughlin, Bodenheimer, 
\& Adams 2004), and the disk gas is likely to have been dissipated
in $\sim$ 3 Myr, well before such a core could accrete a significant gaseous 
envelope (Bally et al. 1998; Bric\~eno et al. 2001; Haisch, Lada, \& Lada 
2001; Eisner \& Carpenter 2003).

Microlensing surveys have also found evidence for two gas giant planets 
with masses of $\sim 1.5 M_{Jup}$ orbiting M dwarfs (Bond et al. 2004; 
Udalski et al. 2005). Because the microlensing signal produced by gas giant 
planets is considerably stronger than that produced by super-Earths,
these four detections suggest that given the limited signal-to-noise
ratios of the ground-based photometry of microlensing events, 
super-Earths must be signficantly more frequent companions to
M dwarfs than gas giants. Thus these four detections have been taken 
as supportive of core accretion's inability to form gas giants around 
low-mass stars (Beaulieu et al. 2006).

Radial velocity surveys were the first to discover the existence
of gas giants and super-Earths around M dwarf stars. The M dwarf
Gl 876 has an outer pair of gas giant planets as well as an
inner super-Earth (Rivera et al. 2005). The M dwarfs Gl 436 and Gl 581
also appear to be orbited by super-Earths on short-period orbits
(Butler et al. 2004; Bonfils et al. 2005). M dwarf planet surveys 
have only been underway for a few years, but they have already
revealed that the frequency of close-in gas giants around M dwarfs appears 
to be lower than that around F, G, and K dwarfs (Endl et al. 2003, 2006).
With time, these surveys will determine the frequency of longer period
gas giants orbiting M dwarfs. Given the results to date of the radial
velocity and microlensing surveys, however, there is a clear need to 
explain the formation of both gas giants and super-Earths around 
M dwarf stars.

We show here that disk instability can explain the formation 
of the gas giant planets orbiting M dwarfs found by both microlensing
and radial velocity, as well as the super-Earths found by microlensing.
We begin by presenting more details about a disk instability
model (5CH) from Boss (2006a) that showed the possibility 
of forming gas giant protoplanets about M dwarfs, and then develop
the reasoning behind the scenario for super-Earth formation
by photoevaporation of gaseous protoplanets orbiting M dwarfs.

\section{Numerical Methods and Initial Conditions}

 Model 5CH (Boss 2006a) was calculated with a finite volume code
that solves the three dimensional equations of hydrodynamics and
radiative transfer, as well as the Poisson equation for the gravitational 
potential. The code is second-order-accurate in both space and time 
(Boss \& Myhill 1992) and has been used and discussed extensively in 
previous disk instability studies (e.g., Boss 2003, 2005, 2006a).

 The equations are solved on a spherical coordinate grid with 
$N_r = 101$, $N_\theta = 23$ in $\pi/2 \ge \theta \ge 0$, 
and $N_\phi = 512$. The radial grid extends from 
4 AU to 20 AU with a uniform spacing of $\Delta r = 0.16$ AU.
The $\theta$ grid is compressed toward the midplane in order to ensure 
adequate vertical resolution ($\Delta \theta = 0.3^o$ at the midplane). 
The $\phi$ grid is uniformly spaced to prevent any azimuthal bias. 
The central protostar wobbles in response to the growth of 
disk nonaxisymmetry, preserving the location of the center 
of mass of the star and disk system. The number of terms in the 
spherical harmonic expansion for the gravitational potential of the disk
is $N_{Ylm} = 48$. 

 Model 5CH calculated the evolution of a $0.5 M_\odot$ 
protostar surrounded by a protoplanetary disk with a mass of 0.065 
$M_\odot$ between 4 AU and 20 AU. The initial protoplanetary 
disk structure was based on an approximate vertical density 
distribution (Boss 1993). The initial disk temperatures were 
derived from the models of Boss (1995), with midplane temperatures of 300 K 
at 4 AU, decreasing monotonically outward to a distance of $\sim 6.7$ AU, 
where they were assumed to become uniform at an outer disk temperature 
of $T_o = 50$ K. The outer disk was then initially marginally
gravitationally unstable in terms of the gravitational stability
parameter $Q$, with an initial minimum value of $Q = 1.5$.

\section{Gaseous Protoplanet Formation}

 Figures 1 and 2 show that model 5CH formed a number of 
clearly defined clumps after 215 years of disk evolution 
around a protostar with a mass of $0.5 M_\odot$. The clumps
grow inside spiral arms, which form in the innermost, marginally
gravitationally unstable regions of the disk ($\sim$ 8 AU), 
because of the combination of relatively low $Q$ and short 
orbital periods there (see Figure 1 of Boss 2006a). This is
largely a result of the initial temperature profile assumed 
for the disk midplane (Boss 1995). Such a profile is to be expected
for a protoplanetary disk in a region of low-mass star formation,
or in any star-forming region prior to the formation of the first 
high-mass stars.

 Table 1 lists the maximum densities in the four clumps evident
in Figures 1 and 2, along with the clump masses $M_c$ in units of
the Jupiter mass $M_{Jup}$, the Jeans mass $M_J$ at the average
density and temperature of each clump, and the instantaneous
values of the orbital semimajor axis and orbital eccentricity of
each clump. In Figure 1, the first clump is located at
11 o'clock, the second at 3 o'clock, the third at 7 o'clock, 
and the fourth at 8 o'clock. Table 1 shows that each clump has a 
mass well in excess of the local Jeans mass, showing that these 
clumps are gravitationally bound. Their effective spherical
radii are comparable to the critical tidal radii at their
orbital distances, implying stability against tidal disruption
by the protostar's tidal forces as well.
 
 Because of the fixed nature of the numerical grid, the code
is not able to provide the locally enhanced spatial resolution
that these high density clumps require for their further
evolution to be calculated correctly. While clump densities
and lifetimes can be increased as the numerical spatial
resolution is increased (Boss 2005), with a fixed grid code
eventually the clumps are sheared apart. Meanwhile, new
clumps continue to form and take their place. Calculations where
the dense clumps are replaced by virtual protoplanets suggest
that the protoplanets should be able to orbit stably for
an indefinite period of time, even as the marginally gravitationally
unstable disk continues to transport mass inward to the
central protostar (Boss 2005). Similarly, SPH code calculations
with a locally-defined smoothing length by Mayer et al. (2002)
have shown that dense clumps should be able to survive their
subsequent orbital evolution, though mergers, scatterings, and 
significant orbital evolution (both inward and outward) are to be
expected when multiple clumps form, as in model 5CH. 

 While we cannot therefore predict the final outcome of model 5CH
with any degree of certainty, the model suggests that
one or more gas giant protoplanets with masses on the order of one 
to a few Jupiter masses could form by disk instability around an M 
dwarf star, with initial semimajor axes on the order of $\sim 10$ AU.
In addition to mutual scattering events, Type II migration during the
disk's lifetime could force gas giants to migrate closer to their
stars. It should be noted that model 5CH was not designed to
attempt to form clumps {\it in situ} at the $\sim$ 2.5 AU orbital
separation at which the microlensing techinique is most sensitive, 
but that with minor changes in the assumed initial disk profiles
(i.e., a cooler inner disk), such an outcome would be likely.
 
\section{Super-Earth Formation}

 If the M dwarf star and disk system represented by model 5CH 
had formed in a region of low-mass star formation like Taurus
or Ophiuchus, the M dwarf would be expected to be accompanied by
one or more $\sim 1 M_{Jup}$ gas giant planets orbiting at
distances of $\sim$ 10 AU or less, possibly explaining the
two gas giant planet microlensing detections (Bond et al. 2004; 
Udalski et al. 2005). However, most stars are formed in
regions of high-mass star formation (Lada \& Lada 2003), 
similar to the Orion and Eta Carina nebulae, where protoplanetary
disks are subjected to a withering flux of FUV/EUV radiation from the 
nearby O stars (e.g., Bally et al. 1998). In the Eta Carina nebula, FUV/EUV 
fluxes are a factor of $\sim 100$ times higher than in Orion, yet 
protoplanetary disks are as commonplace in Carina as in Orion 
(Smith, Bally, \& Morse 2003). Armitage (2000) found that the EUV flux 
alone in an Orion-like cluster was sufficient to photoevaporate
gaseous disks within $\sim$ 1 Myr around stars within 0.3 pc of the
massive stars. In larger clusters like Eta Carina, similarly
rapid photoevaporation would occur for disks within 3 pc of the O stars.

 Boss, Wetherill, \& Haghighipour (2002) suggested that the Solar System 
was formed in a region of future high-mass star formation, such that after
the massive stars formed, their FUV/EUV radiation was able to photoevaporate 
away not only the outer regions of the solar nebula, but also the gas
envelopes of the two outermost gas giant protoplanets formed
by disk instability (Boss 2003), stripping these two gaseous 
protoplanets down to rock/ice cores with only minor gaseous envelopes, 
i.e., turning them into the two ice giant planets, Uranus and Neptune.

 One critical component of this scenario for ice giant planet formation
is for the heavy elements to coagulate into dust grains and sediment
down to the centers of the protoplanets faster than the protoplanets
contract to planetary densities. Boss (1998) estimated that core
formation in gaseous protoplanets would occur in $\sim 10^3$ yr,
with a $1 M_{Jup}$ protoplanet being able to form at most a
$6 M_\oplus$ rock/ice core. Helled, Kovetz, \& Podolak (2006)
performed a more detailed analysis, and confirmed that dust grains
would settle down to form a central core in $\sim 10^3$ yr in
a non-convecting protoplanet. When the effects of convective
turbulence were included, the grains grew faster and reached
the core in $\sim 30$ yr. Helled, Podolak, \& Kovetz (2006)
found that a $1 M_{Jup}$ protoplanet requires $\sim 3 \times 10^5$ yr 
to contract from a radius of $\sim 0.5$ AU to $\sim 0.1$ AU.
During this slow contraction phase, the protoplanet is able
to accrete a significant number of km-sized planetesimals
by gas drag capture in the protoplanet's outer layers. These
planetesimals will either be added to the solid core or will
remain in the protoplanet's envelope, but in either case the
protoplanet will be highly enriched in heavy elements compared
to the solar composition (Helled, Podolak, \& Kovetz 2006).
While much remains to be determined, it seems likely that
gas giants with core masses and envelope enrichments similar
to those of Jupiter and Saturn (Saumon \& Guillot 2004) can
be formed by disk instability.

 The key factor for whether a gaseous protoplanet becomes a gas giant 
or an ice giant is the critical orbital radius $r_e$ outside of which
photoevaporation can remove the disk gas, and hence the protoplanetary
envelope gas. A gravitational radius $r_g$ can be defined to be the 
orbital radius where the sound speed of the UV-heated gas equals the 
gravitational escape speed from the protostar

$$ r_g \approx {G M_p \over c_s^2},$$

\noindent
where $M_p$ is the protostar mass and $c_s \approx 10$ km s$^{-1}$ for gas 
heated by EUV radiation and $c_s \approx 3$ km s$^{-1}$ for FUV radiation.
Depending on the details of the photoevaporation model, $r_e$ is
expected to be smaller than $r_g$, with $r_e \sim 0.5 r_g$
(Johnstone et al. 1998) or even smaller (Adams et al. 2004).
With $r_e \sim 0.5 r_g$, for a G dwarf star like the Sun, 
$r_e \sim 50$ AU for FUV radiation and $r_e \sim 5$ AU for EUV radiation.
Adams et al. (2004) argue that these values of $r_e$ could be
smaller by factors of as large as 5, i.e., to as small as $\sim 10$ AU 
to $\sim 1$ AU for FUV and EUV, respectively. While the exact
value of $r_e$ for any protostar will depend on the relative
amounts of FUV and EUV received, we can calibrate $r_e$ by
noting that if photoevaporation was involved in the formation
of the Solar System's giant planets (Boss et al. 2002), then
evidently $r_e < 9$ AU, in order to result in Saturn's bulk
composition (assuming little orbital migration of Saturn after its
formation at $\sim$ 9 AU). 
 
 These critical orbital radii depend linearly on the mass of the protostar.
For an M dwarf star with 1/3 the mass of the Sun, $r_e$ would
then be expected to be $\sim$ 3 AU or less, assuming the M dwarf protostar
was exposed to the same FUV/EUV environment as the solar nebula.
Gaseous protoplanets orbiting at this distance or beyond would lose the
bulk of their hydrogen gas by photoevaporation, leaving
a protoplanet composed primarily of the residual heavy elements
(helium gas would also be lost by entrainment in the photoevaporative 
hydrogen flow). As a result, the final planet would be composed almost
exclusively of the core and envelope heavy elements.
The models of Helled, Podolak, \& Kovetz (2006) suggest that a
$\sim 1 M_{Jup}$ protoplanet formed by disk instability could
accrete as much as $\sim 30 M_\oplus$ of km-sized planetesimals 
from the protoplanet's feeding zone, using the same assumptions
as are used in core accretion models (e.g., Pollack et al. 1996).
Planetesimal scattering was neglected in these models, so this
estimate appears to be a rough upper bound. [Cometesimal scattering 
by the giant planets is thought to be the source of the Oort Cloud
comets and of the scattered disk component of the Kuiper Belt.]

 Model 5CH produced several $\sim 1 M_{Jup}$ gaseous protoplanets
orbiting an M dwarf at $\sim 10$ AU. If this system formed in
a region of future intense FUV/EUV radiation, these protoplanets would
be stripped down to cores composed of heavy elements, with
masses no larger than $\sim 30 M_\oplus$. Assuming that mutual
scattering and/or Type II migration had resulted in one of
these planets ending up on a $\sim 2.5$ AU orbit, it could then
be detected as a super-Earth by microlensing surveys.
 
\section{Conclusions}

 Boss (2006b) pointed out that if the Orion-Carina scenario of
Boss et al. (2002) was responsible for the formation of the
Solar System's ice giant planets, then a possible test of
this combination of disk instability and photoevaporative losses
would be to see if the dividing line ($r_e$) between gas
giants and ice giants depends on the stellar mass: for lower
mass stars, this critical radius should decrease proportionately.
It remains for this prediction to be tested by future extrasolar
planet searches.

 Because most M dwarf stars are expected to have been subjected to
a high FUV/EUV radiation environment (Lada \& Lada 2003), 
most M dwarf planets found by microlensing surveys would 
be expected to be super-Earths rather than gas giants, as
seems to be the case so far, albeit based on a small sample of 
only four detections to date. Future microlensing detections
will be important to determine if this provisional interpretation 
is correct.

 The lower frequency of gas giant planets on short-period orbits
around M dwarfs compared to F, G, and K dwarfs (Endl et al. 2003, 
2006) may be a result of faster inward orbital migration around
the more massive dwarfs. The well-known correlation of the presence
of gas giants with the metallicity of the host star appears to
be strongest for short-period planets (Sozzetti 2004), consistent
with the expectation that Type II inward migration will be faster
in metal-rich disks (Livio \& Pringle 2003) and could thus result in a 
higher frequency of short-period gas giants (Boss 2005). The rate of 
Type II migration depends on the disk's kinematic viscosity $\nu$, and in 
standard viscous accretion disk theory (e.g., Ruden \& Pollack 1991) 
$\nu = \alpha c_s h$, where $\alpha$ is a free parameter, $c_s$ is 
the sound speed, and $h$ is the disk thickness. M dwarfs will have
cooler disks than G dwarfs (Boss 1995) because of their shallower
gravitational potential wells and their (presumed) proportionately 
lower disk masses and hence smaller optical depths. As a result,
M dwarf disks should be thinner and have smaller sound speeds than
G dwarf disks, leading to smaller values of $\nu$ and hence longer 
Type II migration times. M dwarf planets should thus be less
likely to undergo significant Type II migration prior to removal
of their disk gas.

 There is still an important role to play for the first step of 
the core accretion process in forming planets around M dwarfs.
One of the short-period super-Earths found by the radial velocity 
surveys of M dwarfs (Gl 876 -- Rivera et al. 2005) is known to be 
accompanied by two outer gas giant planets, implying that the super-Earth
formed interior to the gas giants. The scenario presented in this
paper would not be able to explain the formation of the Gl 876 system, 
unless the planets were able to interchange their radial ordering, 
which seems unlikely. Hence the Gl 876 super-Earth is likely to have been 
formed interior to its gas giants by the same collisional accumulation 
process that led to the formation of the terrestrial planets in our Solar 
System (e.g., Wetherill 1996). Taken as a whole, M dwarfs thus appear to 
present strong evidence for the formation of the same three classes 
of planets found in our Solar System: inner terrestrial planets
formed by collisional accumulation, and outer gas giants or rock/ice 
giants (super-Earths) formed by disk instability, either in 
the absence of, or in the presence of, strong fluxes of FUV/EUV 
radiation, respectively.

 I thank Michael Endl for discussions about M dwarf planets, 
the referee for several improvements to the manuscript, 
and Sandy Keiser for her computer systems expertise. This research 
was supported in part by NASA Planetary Geology and Geophysics
grant NNG05GH30G and by NASA Astrobiology Institute grant NCC2-1056.
The calculations were performed on the Carnegie Alpha Cluster, 
the purchase of which was partially supported by NSF Major Research
Instrumentation grant MRI-9976645.

\clearpage
\begin{deluxetable}{cccccc}
\tablecaption{Clump properties for model 5CH at 215 yr. \label{tbl-1}}
\tablehead{\colhead{Clump} & 
\colhead{$\rho_{max}$ (g cm$^{-3}$)} & 
\colhead{$M_c/M_{Jup}$} & 
\colhead{$M_J/M_{Jup}$} & 
\colhead{$a$ (AU)} & 
\colhead{$e$} }
\startdata
1 & $6.2 \times 10^{-10}$ & 1.0 & 0.74 & 8.7 & 0.040 \nl
2 & $2.7 \times 10^{-9}$ & 1.1 & 0.51 & 7.9 & 0.10 \nl
3 & $1.4 \times 10^{-9}$ & 1.3 & 0.57 & 9.0 & 0.093 \nl
4 & $1.6 \times 10^{-9}$ & 1.4 & 0.72 & 8.0 & 0.011 \nl
\enddata
\end{deluxetable}
\clearpage

\begin{figure}
\vspace{-2.0in}
\plotone{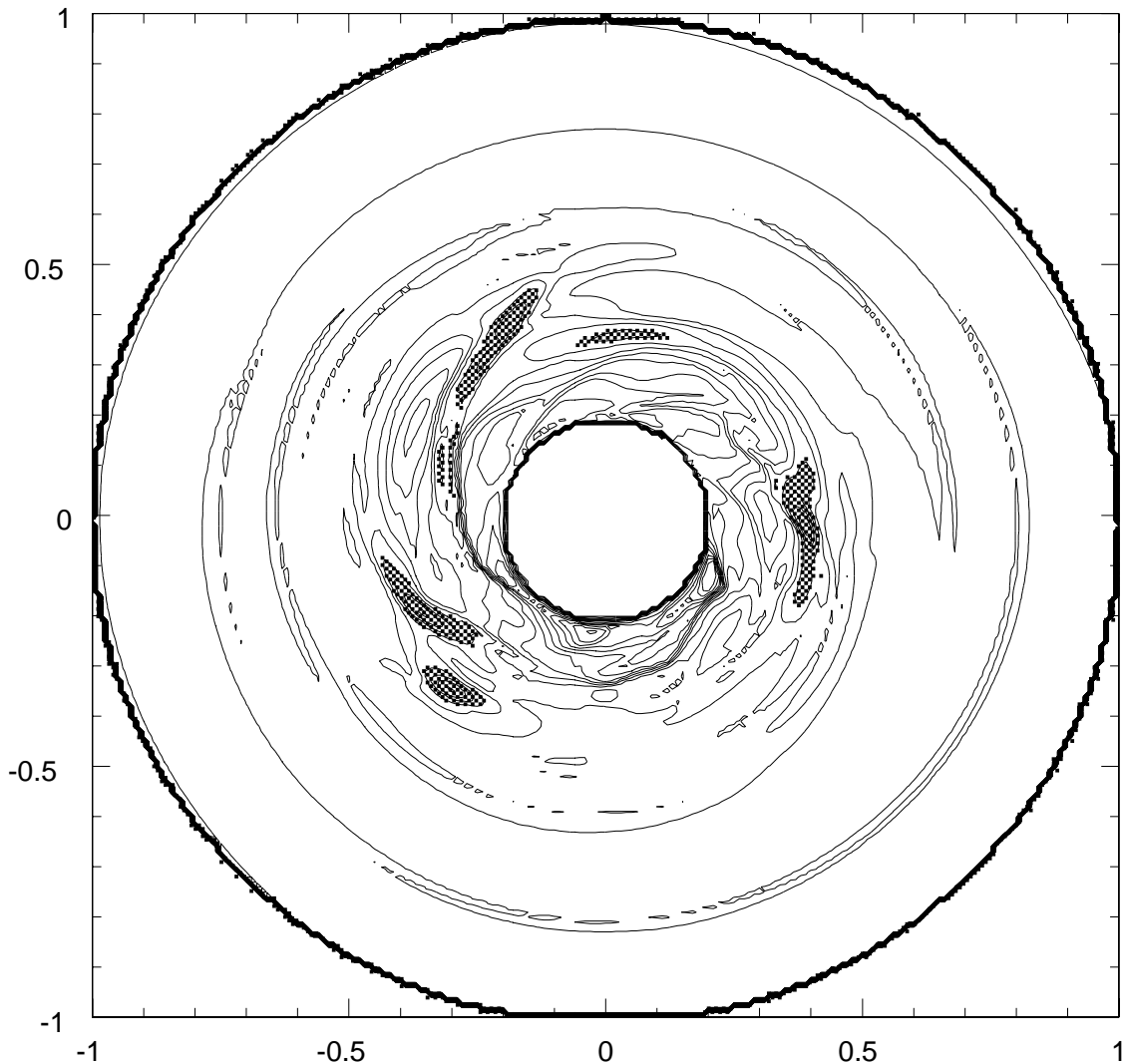}
\caption{Equatorial density contours for model 5CH after 215 yrs of 
evolution. The entire disk is shown, with an outer radius of 20 AU and an 
inner radius of 4 AU, through which mass accretes onto the central protostar. 
Hashed regions denote spiral arms and clumps with densities higher than 
$10^{-10}$ g cm$^{-3}$. Density contours represent factors of two change 
in density. The four clumps described in Table 1 are numbered sequentially 
in counterclockwise order as they appear in this Figure, starting with
clump \#1 at 11 o'clock.}
\end{figure}

\begin{figure}
\vspace{-2.0in}
\plotone{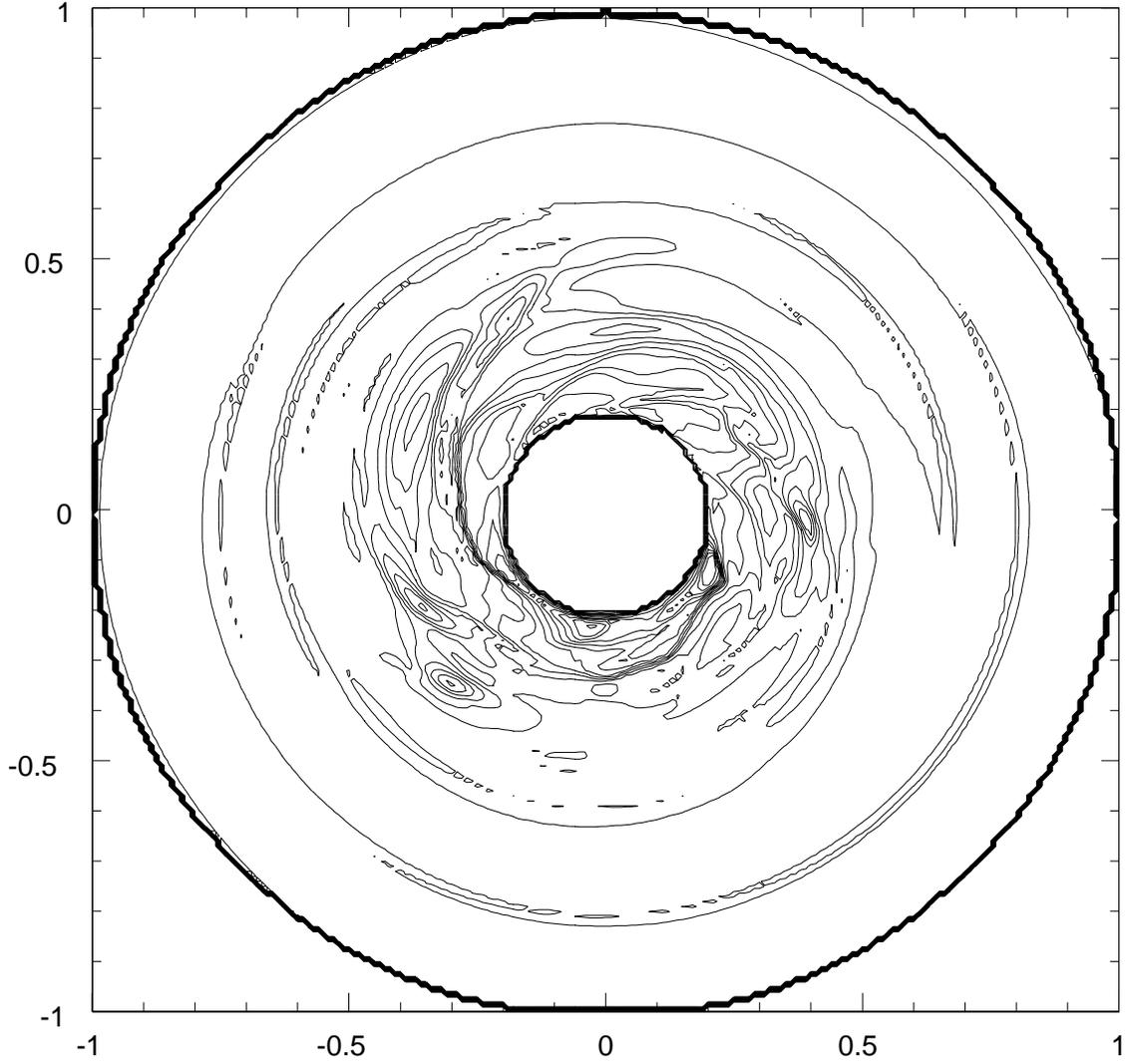}
\caption{Same as Figure 1, but with cross-hatching removed to
reveal the structure of the density contours in the densest regions.}
\end{figure}

\end{document}